# Functional Degradation and Self-enhanced Elastocaloric Cooling Performance of NiTi Tubes under Cyclic Compression

Dingshan Liang[1, 2], Peng Hua[1], Junyu Chen[3], Fuzeng Ren[2,*], Qingping Sun[1,*]

[1]Department of Mechanical and Aerospace Engineering, The Hong Kong University of Science and Technology, Clear Water Bay, Kowloon, Hong Kong, China

[2]Department of Materials Science and Engineering, Southern University of Science and Technology, Shenzhen, Guangdong, China

[3]Department of Engineering Mechanics, School of Civil Engineering, Wuhan University, Wuhan, Hubei, China

Corresponding Author. E-mail: renfz@sustech.edu.cn; meqpsun@ust.hk

## Abstract

Superelastic NiTi tubes are promising candidates for eco-friendly elastocaloric cooling, but their cyclic stability suffers severely from functional degradation. Herein, we investigate the functional degradation of nanocrystalline NiTi tubes via in-depth analysis and find out that it is beneficial to elastocaloric cooling performance. The results show that the functional degradation accompanies with progressive accumulation of residual strain and significant reduction in both hysteresis loop area (*D*) and forward transformation stress ($\sigma_f^{tr}$). The accumulation of residual strain arises from phase-transition-induced dislocations and dislocation-pinned residual martensite. The former separates the original austenite grains to much smaller nanodomains (equivalent to grain size effect) and contributes to the strain hardening during phase transition, leading to the significant reduction of *D*. The latter induces compressive residual stress in the austenite phase and thus gives rise to the evolutive reduction of $\sigma_f^{tr}$. Consequently, the material coefficient of performance (*COP$_{mater}$*) was self-enhanced for 40~104 %. The beneficial effect is mainly because of the cyclically-decreased *D*. The study might provide a processing route to tailor *COP$_{mater}$* and stabilize the mechanical response of NiTi by cyclic compression.

***Keywords:*** Elastocaloric cooling; Cyclic response; NiTi; Shape memory alloy (SMA); Martensitic phase transformation.

# 1. Introduction

In recent years, solid-state cooling using green elastocaloric materials has emerged as an active field of research [1-3]. Superelastic NiTi shape memory alloy (SMA) has been employed as a promising core element in the bulk refrigeration prototypes, due to its large latent heat under stress-induced phase transition (PT) [4-7]. An external applied stress is required to convert the solid-state austenite phase (B2) to solid-state martensite phase (B19′) and the latent heat is released to the surroundings, leading to a temperature increase. With the stress being removed, B19′ transforms to B2 reversely and the same amount of latent heat is absorbed from the surroundings, leading to a temperature drop [8, 9]. Such reversible PT of NiTi can be utilized in solid-state refrigerators based on the elastocaloric effect [10-15]. Of all the stress-loading modes, cyclic compression of bulk superelastic NiTi tubes is favored for elastocaloric cooling, because the tube configuration has the following advantages: (1) much improved fatigue resistance via suppression of crack initiation and growth (up to 78 million ), (2) macroscopically homogeneous stress and strain, and (3) efficient heat transfer from the material to the heat transfer fluid [16-20].

However, conventional bulk NiTi SMA suffers from severe functional degradation, which is manifested as the fast accumulation of residual strain ($\varepsilon^{res}$) and decreases in the hysteresis loop area (*D*) and the forward transformation stress ($\sigma_f^{tr}$) in the first several hundreds of cycles [18, 21-25]. The macroscopic functional degradation in superelasticity strongly affects the functional performance of NiTi SMA in damping structure [26] and actuators [27]. It is well known that such functional degradation originates from the lattice mismatch between B2 and B19′[28, 29]. Under external loading, the mismatch at the atomic scale results in high stress concentration at the habit planes (interface of B2 and B19′) during PT and thus leads to the formation of dislocations [30-32]. Subjected to cyclic loading, the habit planes swipe through the grain interior and even across grain boundaries. Owning to the PT-induced dislocations block the movement of habit planes and suppress the reverse transition of martensite, irreversible residual martensite is constrained at the microscale and the resultant residual stresses are induced in both austenite and martensite phases [23, 33-35]. Such microstructural changes of NiTi under cyclic PT are responsible for the macroscopic functional degradation, which can be reflected from the evolution of the isothermal stress-strain ($\sigma\sim\varepsilon$) responses. Concerning the well-studied mechanisms of functional degradation were mainly characterized under cyclic tension, the fundamental mechanisms under cyclic compression of bulk NiTi require further investigation. Therefore, it is crucial to probe the functional degradation of NiTi tube under cyclic compression, which has not been well-explored, to correlate the microstructural evolution to the macroscopic responses.

Besides, the effects of functional degradation on the elastocaloric cooling performance of NiTi SMA has not been systematically studied. The reduction in the forward transformation stress can increase the transformation strain and temperature drop of NiTi under partial PT [5, 16]. And the decrease in the hysteresis loop area can decrease the amount of input work ($w$) for elastocaloric cooling [3, 36]. Therefore, the functional degradation of SMAs might be beneficial to the elastocaloric cooling performance. It is essential to unravel the relation between functional degradation and elastocaloric cooling, so as to find the optimal working condition of NiTi.

In this work, cyclic compression of NiTi tubes under stress from partial to full PT were performed to evaluate the functional degradation and its effect on the elastocaloric cooling performance. Residual stress and volume fraction of B2 and B19′ were measured from X-ray diffraction (XRD). Microstructural evolution was examined by transmission electron microscopy (TEM). Subsequently, the roles of PT-induced dislocations and residual martensite on the evolution of mechanical responses were analyzed. Finally, the effects of functional degradation on the elastocaloric cooling performance were evaluated. The study provides possible routes to optimize the functional performance of NiTi tubes via cyclic compression.

## 2. Materials and methods

*2.1 Materials and characterization*

Bulk NiTi tubes (Vascotube, Germany) had a wall thickness of 0.55 mm with an outer diameter of 5 mm and contained 50.57 at.% Ti as determined by energy dispersive X-ray spectroscopy (EDX, Oxford Instruments, UK). The austenite finish temperature ($A_f$) was determined to be 21.6 ºC by differential scanning calorimeter (DSC; TA Q1000, USA) ranging from − 80 ºC to 100 ºC at a rate of 10 ºC/min (Fig. S1). The phase composition was identified using X-ray diffraction (XRD, Rigaku Smartlab-9 kW, Japan) in a 2$\theta$ range from 35º to 50º at a step size of 0.01° and a scan rate of 10° min$^{-1}$. The *in situ* XRD was conducted at a heating/cooling rate of 10 °C/min and held for 20 minutes at each tested temperature. The quantitative analysis were evaluated via Rietveld refinement in MAUD. The microstructure was examined by TEM (200 kV, JEOL 2010, Japan and 300 kV, FEI Tecnai G2 F30, USA). The TEM thin foils were lifted out from the middle section of the tubes using focus ion beam system (FIB; FEI Helios G4, USA).

*2.2 Mechanical testing set up*

The mechanical tests were performed on an MTS machine (Landmark 370.10, USA) with a specially designed compression fixture for NiTi tubes (Fig. S2a). The temperature of clamped tubes under MTS frame is ~ 30 ºC, which is above $A_f$. The *in-situ* temperature profile was captured by an infrared camera (FLIR SC7700M, USA). The NiTi tubes were cut into 22.5 mm (with an effective height of 12.5 mm; Fig. S2b) in height by electrical discharge machining. The radial direction, axial direction and tangential direction were denoted as RD, AD and TD. Theoretical estimation showed that the tubes would not buckle under the present testing conditions (details can be found in Part 3 of Supplementary data). And it has also been confirmed by the experimental observations during cyclic compression.

To determine the strain of the tubes, empty tests were run first without tubes to determine the force-displacement relation of the compression fixture and the MTS frame. All the data obtained from the MTS system were proceeded to eliminate the displacement of the compression fixture and the MTS frame for a relatively accurate displacement of tested tubes.

*2.3 Cyclic compression, isothermal test, and COP measurement*

The compressive stress-strain curve (Fig. S3) of the as-received NiTi tubes show that $\sigma_f^{tr}$ is approximately 617 MPa at room temperature. Thus, three maximum applied loads ($\sigma_{max}$) of 800, 1000 and 1200 MPa were selected to study the cyclic response from partial PT to full PT. Cyclic compression were conducted at a frequency of 4 Hz to accumulate cycle number (*N*). It was stopped after certain cycles to perform isothermal compression at a loading rate of 6 N/s and *COP* measurement. The *COP* measurements were conducted for five cycles right after the isothermal compression. The procedure is set as: 1) adiabatic loading to $\sigma_{max}$ in 0.2 s; 2) holding the displacement for 70 s to transfer heat to air ambient; 3) adiabatic unloading to 50 MPa in 0.2 s; 4) holding the displacement for 70 s to transfer heat to air ambient. The obtained last four cycles in the *COP* measurement, including the complete circle of the thermodynamic cycle, were used to analyze the mechanical input work, adiabatic temperature drop ($\Delta T_c$) and *COP$_{mater}$*.

*2.4 Temperature dependence of the forward transformation stress*

Three tubes were loaded to 1200 MPa at a loading rate of 6 N/s on an Instron 5969 testing machine (USA) with a furnace at 25 , 35 and 45 °C. The forward transformation stress and phase transformation strain ($\varepsilon^{tr}$) were obtained from the isothermal strain-stress curve.

*2.5 Removal of residual martensite via low-temperature heating*

A tube was cyclically compressed under $\sigma_{max}$ of 1200 MPa at 4 Hz for $10^4$ cycles and the change of height was recorded to study the quantitative contribution of residual strain. The deformed tube was heated at 100 ºC for 20 minutes and XRD was performed on the surface after it was cooled to room temperature. Then, two cycles of isothermal compression were conducted to make a comparison with the non-heated tube. Finally, the heated tube was stabilized to $\varepsilon^{res}$ = 3.88 % with another $10^4$ cycles. The *in situ* XRD was carried on the tube surface after $2\times10^4$ cycles with a residual strain of 3.88%. The XRD on the RD-TD plane were performed on the as-received, deformed ($\varepsilon^{res}$ = 3.88 %) and heated specimens.

## 3. Results and discussions

*3.1 Basic features of NiTi tubes*

The as-received tubes have nano-laminated gains with an average layer thickness of 78 nm (Fig. 1a). The selected area diffraction pattern (SADP) confirms the presence of B2 austenite at room temperature and the texture is mainly <110>//RD (inset of Fig. 1a). The typical grain size of B2 (110) is about 44 nm (bright in Fig. 1b). At 25 ºC, $\sigma_f^{tr}$ is 613 MPa and $\varepsilon^{tr}$ is 3.1 % (Fig. 1c). $\varepsilon^{res}$ is 0.4 %, 0.51 % and 1.04 % at 25 ºC, 35 ºC and 45 ºC under 1200 MPa, indicating higher temperature coupling with higher $\sigma_f^{tr}$ results in a higher degree of irreversibility during the reverse PT. As reported in the literatures [4, 6, 37, 38], the temperature dependence of $\sigma_f^{tr}$ is linear as fitted in Fig. 1d. The Clausius-Clapeyron slope ($K_f$) is determined to be 12.04 MPa/ºC. Based on the Clausius-Clapeyron equation, the transition entropy change ($\Delta S$) can be evaluated from $\Delta S = -K_f \cdot \varepsilon^{tr}$ [15]. At 25 °C, $\Delta S = -0.37324$ MJ/(m$^3$·°C). Take the density of NiTi as 6500 kg/m$^3$, $\Delta S$ is as large as $-57.4$ J/(kg·°C). Of particular note is that the tube can produce an adiabatic temperature drop of 25 °C during reverse PT under a preloading stress of 800 MPa (see Part 5 in Supplementary data). In short, the NiTi tubes have a large transition entropy change and demonstrate promising cooling potential.

*3.2 Functional degradation of NiTi tubes*

*3.2.1 Evolution of the isothermal σ~ε responses*

The functional degradation of NiTi tubes can be reflected from the evolution of compressive σ~ε responses under 800, 1000 and 1200 MPa. As shown in Fig. 2, the functional degradation displays as the accumulation of $\varepsilon^{res}$ and the reduction in $\sigma_f^{tr}$ and *D*. The isothermal σ~ε curves evolve from the typical plateau type at the beginning to a strain-hardening one as the cycle accumulates. Based on the slope transition, the phase transformation stresses can be determined and the loading stage of the σ~ε curves can be divided as the elastic deformation of B2 ($E_{B2}$), PT (B2→B19′) and elastic deformation of B19′ ($E_{B19'}$). The reverse transformation stress ($\sigma_r^{tr}$) with nonobvious slope transition was determined as the stress at the same strain of $\sigma_f^{tr}$ during unloading.

Under $\sigma_{max}$ of 800 MPa (Fig. 2a), the NiTi tube undergoes partial PT with a $\varepsilon^{tr}$ of 2.5%, and there appears to be no sign of $E_{B19'}$ from the σ~ε curve at *N* = 1. The σ~ε curve changes from a broad PT hysteresis loop to a narrow one when the cycle increases. Meanwhile, $E_{B2}$ couples with PT and more amount of B2 can be transformed to B19′ since $\sigma_f^{tr}$ keeps decreasing. Such evolution can be confirmed by the cyclically-increased temperature change under 4 Hz after 100 cycles (Noted in Fig. 2a, see Part 6 in Supplementary data for detail). The quicker the temperature rises versus time after certain cycles of compression (Fig. S5b-c), which is consistent with the reduction of $\sigma_f^{tr}$.

For tubes under 1000 MPa (Fig. 2b) and 1200 MPa (Fig. 2c), full PT can be achieved since $E_{B19'}$ can be determined from the increase of slope after the PT plateau in the first cycle. Akin to the case under 800 MPa, $\sigma_f^{tr}$ and $D$ decrease dramatically as the cycle increases. Under 1000 MPa, the $\sigma \sim \varepsilon$ curve changes from a large open loop to an enclosed middle-narrow one with sharp ends. As for 1200 MPa, it develops from a huge middle-wide loop to a curved "needle" tilted to the top-right. Such geometrical changes in the hysteresis, accounting for the strain hardening at the stage of $E_{B19'}$, might be attributed to the martensite reorientation [39-41].

The detailed evolutions of $\varepsilon^{res}$, $\frac{d\varepsilon^{res}}{dN}$, $\sigma_f^{tr}$, $\sigma_r^{tr}$, $D$ and $\frac{dD}{dN}$ were plotted versus cycle $N$ in Fig. 3 to study the functional degradation. These evolutions demonstrate linear relationships in the logarithmic scale before saturation. That is, the macroscopic degradation features obey the power-law relationships similar to that at the microscale [29], while the cyclic effect is much faster. The height change of the tubes reveals the residual strain due to cyclic compression (Fig. 3a). Gradually, the residual strains stabilize at 1.34 %, 2.86 % and 3.88 % under $\sigma_{max}$ of 800, 1000 and 1200 MPa after $10^3$ cycles. The accumulation rate of $\varepsilon^{res}$ demonstrates a monotonical manner to decrease with increasing $N$ and becomes 4 orders lower after $4\times10^3$ cycles under $\sigma_{max}$ of 800 MPa, as small as $10^{-7}$ per cycle. It drops to 0 after $10^3$ ($\sigma_{max}$ = 1000 MPa) and $4\times10^3$ ($\sigma_{max}$ = 1000 MPa) cycles, without appearing in the logarithm scale in Fig. 3b. Overall, The residual strain is highly sensitive to $\sigma_{max}$ and higher $\sigma_{max}$ leads to quicker stabilization to a larger saturated magnitude after $10^3$ cycles.

Similarly, $\sigma_f^{tr}$ drops significantly in the first 400 cycles and gradually stabilizes after $10^3$ cycles (Fig. 3c). It can be reduced to as low as 177 MPa ($\sigma_{max}$ = 1000 MPa) and 163 MPa ($\sigma_{max}$ = 1200 MPa) at $N=10^4$, which is only one third of that at $N = 1$. Likewise, $\sigma_r^{tr}$ behaves in a tendency to reduce as $N$ increases (Fig. 3d). Both $\sigma_f^{tr}$ and $\sigma_r^{tr}$ are in a higher level with respect to lower $\sigma_{max}$. The hysteresis loop area, which reflects the hysteresis heat [42], keeps decreasing in power-law relation and reaches to an analogous magnitude below 2 MPa at $N=10^4$. The reduced $D$ is about 5, 6.5, and 8 times smaller than that at $N = 1$ for $\sigma_{max}$ of 800, 1000 and 1200 MPa (Fig. 3e). The decreasing rate of $D$ ($dD/dN$) drops significantly to the scale of $10^{-4}$ MPa/cycle after $4\times10^3$ cycles, which is about 5 orders lower than that at the beginning (Fig. 3f). Overall, higher $\sigma_{max}$ results in quicker stabilization within $10^3$ cycles and larger amount of degradation.

### 3.2.2 Cyclically-induced residual martensite

To seek the origins of the functional degradation, we performed XRD characterization on the middle section of tubes after $10^4$ cycles and made comparisons with the as-received tube. The diffraction profiles were taken under the same conditions and indexed by B2 (JCPDS No. 65-0917) and B19′(JCPDS No. 65-0145). Such comparisons can provide solid evidence of the phase evolution due to cyclic compression.

Compared with the as-received tube, the diffraction peak of B2 (110) of the cyclically-deformed tubes at $N = 10^4$ show three distinct features with the increase of $\sigma_{max}$: (i) shift to a larger $2\theta$; (ii) pounced peak broadening; and (iii) much weaker intensity (Fig. 4a). These observations suggest the formation of lattice strain, buildup of dislocations and size reduction of B2. In contrast, the characteristic peaks of B19′ become significantly stronger in the cyclically-deformed tubes (Fig. 4b). Evidenced by the presence of B19′ (012), the residual martensite coexisted with austenite in the cyclically-deformed tubes and higher $\sigma_{max}$ refers to a greater amount of residual martensite.

Based on the shift in the $d$-spacing (Bragg's law: $2dsin\theta = n\lambda$), the uniform lattice strain can be calculated via the following equation [23].

$$\varepsilon_{hkl} = \frac{d_{hkl}^{deformed} - d_{hkl}^{as-received}}{d_{hkl}^{as-received}} \qquad (1)$$

The lattice strain from the indexed planes is plotted in Fig. 4c. It is found that the lattice strain of B2 (110) and B19′ (002) are in an equivalent magnitude but the sign is opposite. It is compressive for B2 (110) while tensile for B19′ (002) along RD, demonstrating a higher degree of lattice strain under higher $\sigma_{max}$. Such distributions of internal strain filed reveal that the residual martensite is capable to coexist with austenite without external stress, similar to the observed results under cyclic tension [23, 43, 44]. Concerning to the compressive lattice strain on B2 (110), the residual stress is estimated based on the elastic modulus of B2 (110) ($E_{110}^{B2} = 74.7$ GPa) under compression [41].

The residual stress of B2 (110) at $N = 10^4$ is higher with respect to higher $\sigma_{max}$ (Fig. 4d). Although this residual stress is along RD (perpendicular to the compression direction), such buildup of residual stress in B2 may be responsible for the progressive change in $\sigma_f^{tr}$.

The full width at half maximum (FWHM) of B2 (110) tends to increase with respect to higher $\sigma_{max}$, suggesting a higher degree of grain size reduction and/or lattice distortion in B2 under higher $\sigma_{max}$ (Fig. 4e). To further understand the effect of residual martensite on the functional degradation, we estimate the volume fraction of B2 and B19′ via Rietveld refinement (Fig. 4f). Detailed refinement analysis can be found in Fig. S6. The results show that the as-received tube contains 2.4 % of martensite, which might be induced by the processing of tube-drawing and the near room temperature $A_f$. The volume fraction of residual martensite increases to be 18.4%, 31.6% and 37.7% for the cyclically-deformed tubes under $\sigma_{max}$ of 800, 1000 and 1200 MPa. It shows that higher $\sigma_{max}$ leads to a greater amount of residual martensite. Considering similar tendency of residual strain, the residual martensite may contribute to the residual strain at the macroscopic scale.

### 3.2.3 Cyclically-deformed microstructure: the role of dislocations

For further observation at microscale, we examined the microstructure of the tube after $10^4$ cycles of compression under $\sigma_{max}$ = 1200 MPa, which has the highest level of macroscopic residual strain. As shown in Fig. 5a, the microstructure of cyclically-deformed tube is significantly different from the original nano-laminated structure, and residual martensite (B19′) is found from the SAED pattern (inset of Fig. 5a). It is notable that the original texture shows very intense diffraction spots of B2 (110) in parallel to RD (Fig. 1a). However, the texture of deformed tube demonstrates a main manner of B2 (110) in parallel to AD, which is the loading direction. The texture of B2 phase after $10^4$ cycles of PT might be self-accommodated to the preferred orientation [45, 46].

The microstructural changes of B2 are mainly size reduction and dislocation slip. After $10^4$ cycles of compression, the nano domains of B2 (110) are partitioned to ~ 9.5 nm by dislocation walls and distributed inside a one-hundred-nanometer B2 realm (Fig. 5b). Such size reduction, equivalent to grain size refinement, is in good agreement with the broadening in FWHM of B2 (110) (Fig. 4e). Note that smaller grain size has been proved to result in smaller hysteresis loop area of NiTi [36, 47, 48]. The significantly reduced size of B2 (110) in the cyclically-deformed tube may be responsible for the greatly decreased hysteresis loop area. Fig. 5c-d show the high-resolution TEM image and the selected area inverse fast Fourier transformation (IFFT) image of B2 austenite. Substantial intragranular lattice distortion is observed of B2 (Fig. 5c). Dislocation slips with Burgers vector of **b** = 0.298 ($\bar{1}$10) are found in a B2 grain (Fig. 5d). These imply that the compressive PT induce dislocation slip and lattice distortion in B2 austenite phase.

Severe lattice distortion of the residual martensite is observed in the high resolution TEM image (Fig. 5e). The lattice scale mismatch is found next to the residual martensite grain with a size of 20 nm. From the IFFT image (Fig. 5f) of the selected area in Fig.

5e, a high density of dislocation blocks and surrounds the grain of B19′. The dislocation density in the selected area is as high as $3.18\times10^{12}$ cm$^{-2}$, almost a half of that at the interface between crystalline and amorphous phase in cold-rolled NiTi [49]. The dislocation slip in B2 and dislocation-pinned martensite may be the main causes to the macroscopic residual strain, similar to the cyclic compression of NiTi micro pillar [29]. In addition, the surface roughness of the cyclically-deformed tubes slightly increases, which may be a result of PT-induced plastic deformation on the surface (Fig. S8).

Such lattice distortion inside a residual martensite grain reveals that the internal strain field plays an important role in the cyclically-deformed NiTi [50]. Regarding a high dislocation density and severe lattice dilation, the shear modulus would be softened according to the Grüneisen relation [49, 51]. That is, the PT (B2→B19′) induced by shear deformation can be triggered easily. It is notable that the B2 grains are partitioned to tens of nanometers by dislocation walls under such high degree of lattice distortion. The size reduction of B2 indicates that the reversible PT could have occurred at lower energy barrier due to the continuous PT with habit planes moving forth and back within reduced domains pinned by dislocations, instead of the habit planes migrating through large grain and even grain boundaries [47]. Also, the dislocation-pinned residual martensite can serve as the nucleation origin of PT, without overcoming high energy barriers for B2 to initially transform to B19′. And thus the transformation stress is reduced during functional degradation. The gradual stabilization of NiTi tubes evolves from the discontinuous first-order PT at the beginning to continuous PT under high internal stress field coupled with dislocation-blocked martensite. Such continuous PT in can permit SMA with small dissipation energy, leading to a small hysteresis loop area [36, 52, 53]. The isothermal $\sigma$~$\varepsilon$ curves also reflect the evolutionary transformation mode from first-order PT (segregation of $E_{B2}$ and PT) to continuous PT (coupling of $E_{B2}$ and PT) (Fig. 2). Consequently, the monotonically reduced hysteresis loop area could be sourced from the reduced size of dislocation-wall-separated B2 grain and the resultant continuous PT.

### *3.2.4 Removal of residual martensite via heating*

Note that the residual martensite can be healed by heating [29, 54]. We further explore the mechanical responses, phase composition and microstructure after removal of residual martensite. Consequently, the effect of residual martensite on the macroscopic $\varepsilon^{res}$, $\sigma_f^{tr}$ and $D$ are analyzed.

We first analyze the variation of mechanical response and corresponding XRD due to heating. As plotted in Fig. 6a, the isothermal $\sigma$~$\varepsilon$ responses after heating (red solid line) represents a variation from stabilized small hysteresis loop (low $\sigma_f^{tr}$, $N = 10^4$) to enlarged open loop (increased $\sigma_f^{tr}$ and partially-recovered $\varepsilon^{res}$). It is notable that the residual strain before heating is 3.8%, and it is in an equivalent magnitude to the tube without heating ($\varepsilon^{res} = 3.88\%$). The main effects of heating are partial recovered $\varepsilon^{res}$, enlarged $D$ and increased $\sigma_f^{tr}$. Evidenced by the XRD pattern after heating, the intensity of the diffraction peaks of residual martensite reduced significantly, suggesting most of the residual martensite is thermally healed back to austenite (Fig. 6b). The characteristic peak of B2 (110) shifts to a smaller $2\theta$ angle compared to the deformed tube and still maintains a slight shift to the right with 0.15º, indicating that the high compressive residual stress in B2 evolves to a lower scale. This result confirms that the change of $\sigma_f^{tr}$ is tightly linked to the compressive residual stress in the B2 phase.

In order to uncover the nature of stabilized microstructure, *in situ* XRD were performed on the deformed tube ($\varepsilon^{res}= 3.88\%$, $N = 2\times10^4$) (Fig. 6c). Another $10^4$ cycles were conducted after two cycles of isothermal compression and the residual strain is stabilized to 3.88 % (equivalent to the non-heating tube at $N = 10^4$). It indicates that the residual martensite can be removed by heating and recreated by cyclic loading, with the obvious change in residual strain. As the temperature cools from 25 °C to − 50 °C, two key features are observed: 1) the intensity of B2 (110) decreases obviously while B19′ (002) increases, especially at − 50 °C (blue line in Fig. 6c); 2) the position of B2 (110) shifts to the right while B19′ (002) to the left. The former suggests that partial B2 is thermally transformed to B19′ and B2 can coexist with B19′ even at − 50 °C, which is well below the original $A_f$. This finding might be able to provide reversible PT of cyclically-deformed NiTi for the elastocaloric cooling at extremely low temperature. The latter reveals that higher degree of internal stress field, as well as lattice distortion reflected from the broadening and asymmetry of the shape of B2 (110), were formed at − 50 °C. Interestingly, such thermally-induced changes are reversible since the diffraction pattern can be recovered when it heated back to 25 °C, demonstrating a two-way shape memory effect. Then the tube was heated to 100 °C and held for 20 minutes, and the qualitative results consist with the comparison of XRD in Fig. 6b.

We then further correlate the residual stress along loading direction to the reduction of $\sigma_f^{tr}$, and hence XRD on the surface of the RD-TD plane were performed. Based on the peak shift, the residual stress of B2 (110) and (211) is compressive in the cyclically-deformed tube (Fig. 7a-b). After heating, such compressive residual stresses change to tensile residual stress and lower degree of compressive residual stress. As $\sigma_f^{tr}$ of the deformed tube is decreased and it partially increases after heating, the compressive residual stress of B2 phase should be responsible for the reduction of $\sigma_f^{tr}$ during functional degradation.

As a complement, we heated the TEM foil (lifted from the cyclically-deformed tube) at 100 ºC for 20 minutes and observed the microstructure. The heated microstructure is similar to the cyclically-deformed one but no diffraction spot of B19′ is observed from the SADP (Fig. 8a). It is basically B2 phase after the heating. Evaluating the average $d$-spacings of B2 (110) from the SADP (Fig. S8), the lattice strain after heating along AD is − 0.006, which is significantly lower than that before heating (− 0.043). Although it is highly localized under TEM, it is consistent with the results of XRD along the loading direction. Regarding the significant reduction of $\sigma_f^{tr}$, it implies that the compression-parallel compressive residual stress accounts mostly for the reduction of $\sigma_f^{tr}$. To conclude, the significant reduction of $\sigma_f^{tr}$ may source from the pile-up of compressive residual stress in the B2 phase.

After confirming the compressive residual stress in B2 lead to the reduction of $\sigma_f^{tr}$, we then further discuss the underlying mechanisms for the shrinkage of $D$. The amount of B2 (110) increases due to the heating (intensity in Fig. 7a). The dark field TEM image further confirms that the grains of B2 (110) with size at tens of nanometers are hierarchically distributed inside the hundred-nanometer domains (Fig. 8b). These results are in good agreements to the comparisons of XRD (Fig. 6-7). Regarding that the hysteresis loop area enlarged while strain hardening remained after heating, the reduction of B2 domain size might be responsible for the strain hardening of the hysteresis loop, partly leading to the shrinkage of the hysteresis loop area. In the other hand, the larger reduction of $\sigma_f^{tr}$ than that of $\sigma_r^{tr}$, which is induced by the internal stress field, results in the decrease of the difference between $\sigma_f^{tr}$ and $\sigma_r^{tr}$ ($\Delta\sigma^{tr}$). The linear relation between hysteresis loop area and $\Delta\sigma^{tr}$ discloses that the smaller $\Delta\sigma^{tr}$ couples with smaller hysteresis loop area (Fig. S9). Therefore, the shrinkage of hysteresis loop area is originated from the reduction of B2 domain size (strain hardening and lower dissipated energy) and internal stress field (reduction of $\sigma_f^{tr}$).

Finally, we estimate the effect of residual martensite on the macroscopic residual strain. After heating, about 1.2% of residual strain is recovered due to the thermal-induced restoration of residual martensite. Referring to the Rietveld refinement, the volume fraction of residual martensite is determined to be 0.6 % after heating. That is, about 37.1 % of residual martensite are thermally-transformed to austenite, recovering 1.2 % of residual strain. Since the transformation strain of the tube is 3.1% at room temperature, the residual strain contributed by residual martensite can be evaluated as : $\varepsilon_{B19'}^{res} = \varepsilon^{tr} \times V_{B19'}$, where $V_{B19'}$ is the volume fraction of contributed martensite. Therefore, $\varepsilon_{B19'}^{res}$ can be evaluated to be 1.15 % and it is well consistent with the recovered residual strain of 1.2%. The calculated result suggests that about 33% of the residual strain is contributed from the residual martensite. The study on the cyclic behavior of NiTi micropillar also confirms that the residual martensite makes certain contribution to the residual strain [29].

*3.2.5 Mechanisms of functional degradation*

The functional degradation in the mechanical response of NiTi tubes is briefly described in Fig. 9a, demonstrating an overall tendency of accumulation of the residual strain, decrease in the forward transformation stress and shrinkage of the hysteresis loop area. As well explained under tension [23], the fundamental origins are the lattice mismatch in the interface between B2 and B19′ and resultant PT-induced dislocations. As presented in Fig. 9b, the fresh tube first undergoes PT from A to B and the forward PT results in the formation of dislocations in the grain and phase boundaries, developing localized stress concentration. Such processes include $E_{B2}$, B2→B19′ and $E_{B19'}$, showing a typical route of first order stress-induced PT. Then during the unloading stage from B to C, lattice strain was produced, and the dislocations block the reverse PT of nanosized martensite domains. After thousands of cycles in compression, the density of dislocation is almost saturated, and the volume fraction of associated residual martensite reaches a stabilized level. The compressive residual stress is mainly in B2 while the tensile residual stress is in B19′. Upon external compressive loading, the B2 domains under high compressive residual stress can transform to B19′ locally even when the external stress is very low. Within the dislocation-wall-separated B2 domains, PT can occur continuously from low stress level to $σ_{max}$, demonstrating the significant strain hardening behavior from D to E [55]. The reversible PT between D and E can occur in a stabilized path due to the high compressive residual stress. It is possible for local PT under low external stress and continuous PT is coupled with $E_{B2}$ within the refined domains.

To conclude, the main mechanisms of functional degradation of NiTi tubes are: 1) PT-induced dislocation slip and dislocation-pinned martensite leads to the accumulation of residual strain; 2) the compressive residual stress in the austenite phase results in the reduction of phase transformation stress; 3) the reduction of B2 domain size and internal stress field contribute to the shrinkage of hysteresis loop area.

*3.3 Self-enhanced elastocaloric cooling performance*

In order to evaluate the elastocaloric performance of NiTi tubes, $COP_{mater}$ and $COP$ of Carnot cycle ($COP_{Carnot}$, the maximum theoretical efficiency that a thermodynamic cycle can achieve [56]) were measured at different stages during the cyclic compression of the tubes. Fig. 10a presents a thermodynamic cycle, including heat transfer with the air ambient and adiabatic loading/unloading to simulate the heating/cooling process. The temperature profile of the NiTi tubes under forward and reverse PT was captured simultaneously (Fig. 10b). The average temperature of the selected area of the tubes shown by the inset in Fig. 10b is exported and plotted against time. The $COP_{mater}$ can be estimated by Eq. 2 from the input work $w$ (=$D$, as shown in Fig. 10a) and the temperature drop $\Delta T_c$ during reverse PT (Fig. 10b).

$$COP_{mater} = \frac{\lambda \cdot \Delta T_c}{w}, \tag{2}$$

where $\lambda$ = 3.225 MJ/(m$^3$·°C) is the specific heat capacity. And the $COP_{Carnot}$ can be estimated by Eq. 3 from the temperature of the heat source ($T_c$) and the heat sink ($T_h$) in a thermodynamic cycle [5, 12, 57, 58].

$$COP_{Carnot} = \frac{T_c}{T_h - T_c}, \qquad (3)$$

where $T_c$ and $T_h$ are in Kelvin. Based on the parameters in Fig. 10a-b, it can be evaluated that: $COP_{mater}$ = 12.65 and $COP_{Carnot}$ = 30.39. The ratio of $COP_{mater}$ to $COP_{Carnot}$ can be seen as the exergetic efficiency evaluating the performance of cooling refrigerant [58, 59].

Based on such evaluation, the evolution of $w$ and $\Delta T_c$ was presented in Fig. 10c-d. Detailed thermo-mechanical cycles and temperature profiles can be found in Part 11 of the Supplementary data. $w$ decreased with the increase of $N$ for full PT under 1000 and 1200 MPa. While for the partial PT under 800 MPa, $w$ increased slightly from 2 ($N$ = 1) to 3.2 MPa ($N = 10^3$) and then decreased to 2.4 MPa ($N = 10^4$). $\Delta T_c$ monotonically decreased as $N$ increased under 1200 MPa, from 15.5 °C to 11.9 °C. Under 1000 MPa, $\Delta T_c$ slightly decreased from 13.8 °C to 13.0 °C within $10^4$ cycles. Under 800 MPa, $\Delta T_c$ initially increased from 6.8 °C to 11.6 °C in the first $10^3$ cycles and then slightly dropped to 9.4 °C at $N = 10^4$. The increase in $\Delta T_c$ might be ascribed to the decrease in transformation stress and the increase in the amount of PT [16].

Given the evolutions of $w$ and $\Delta T_c$, evolution of $COP_{mater}$ and the exergetic efficiency under various $\sigma_{max}$ are quantified in Fig. 10e-f. $COP_{mater}$ demonstrates an overall trend of increasing with the increase in $N$, presenting a self-enhanced behavior. The tube under 800 MPa has the highest $COP_{mater}$ among three stress levels and reaches a maximum value of 12.65 at $N = 10^4$. This advantage maintains throughout the full range of cyclic compression. For the cases of 1000 and 1200 MPa, $COP_{mater}$ increases monotonically and is almost double at $N=10^4$ compared with that at $N = 1$. As for the exergetic efficiency, a general tendency of improvement is found within $10^3$ cycles and then remains above 40% at $N = 10^4$. Under 800 MPa, not only the highest value of $COP_{mater}$ can be achieved but the exergetic efficiency can be improved twice with $10^3$ cycles of compression. The origins of such self-enhancement in elastocaloric performance of the NiTi tube under partial PT are due to the increase of $\Delta T_c$. Under full PT, the significant shrinkage of $D$ has a dominant effect on the increase of $COP_{mater}$. Hence, it is essential to tune the mechanical behaviors via microstructure engineering under suitable stress to minimize the hysteresis loop area (or input work) of NiTi for elastocaloric cooling.

The functional degradation of NiTi generally brings negative effect under real applications, as it demonstrates the unstable cyclic stress-strain responses of NiTi. Nevertheless, the self-enhancement of the elastocaloric cooling performance of NiTi tubes shows that the functional degradation in cyclic $\sigma$-$\varepsilon$ responses of NiTi is beneficial to the cooling performance. Compared with other SMAs of different shapes [5, 12, 57, 60, 61], NiTi tubes demonstrate a good combination of high $COP_{mater}$ and high specific surface area for efficient heat transfer (Fig. 11). $COP_{mater}$ of the NiTi tube under partial PT at $N=10^4$ even exceeds that of TiNiFe thin film [61], indicating that the NiTi tubes under compression are promising alternatives for elastocaloric cooling.

Furthermore, cyclic compression permits NiTi with self-enhanced cooling performance from partial to full PT. It is also a strategy to create a stable nanostructure of NiTi with sub-grain defects of dislocation walls and dislocation-pinned residual martensite. The induced defects can significantly decrease the hysteresis loop area and increase $COP_{mater}$, presenting an effective and simple process to optimize the cooling performance. Introducing defects in NiTi via cyclic compression might be a promising way to fabricate solid-state green refrigerants with tunable $COP_{mater}$ and high cyclic stability. Besides the functional degradation, the study of structural fatigue, optimizations of heat transfer conditions and geometric parameters is under investigation and will be addressed in our future work.

## 4. Conclusions

Functional degradation and elastocaloric cooling performance of superelastic NiTi tubes are explored under cyclic compressive PT. The following conclusions can be drawn from this study:

1) The functional degradation of NiTi tubes under cyclic compression is caused by the PT-induced plasticity, contributed from dislocation slip and residual martensite. The corresponding compressive residual stress in the austenite phase leads to the reduction in forward and reverse PT stress. The dislocation wall partition inside the austenite grains and the internal stress field lead to the shrinkage of hysteresis loop area.

2) The elastocaloric cooling performance of the NiTi tubes under cyclic compression is self-enhanced with the increase in cycle number, because of the functional degradation. It is found that the partial PT of NiTi tubes under cyclic compression is more favorable than full PT for elastocaloric cooling, due to the increases of $\Delta T_c$ and $COP_{mater}$.

3) The cyclically-deformed NiTi tube ($10^4$ cycles of compression under 800 MPa) shows a high $COP_{mater}$ of 12.65. The effective heat transfer structure and high $COP_{mater}$ make NiTi tubes promising materials for green solid-state cooling technology.

4) Introducing defects of dislocations and residual martensite in NiTi via cyclic compression might be a helpful processing method to improve cyclic stability and $COP_{mater}$.

## Acknowledgements

This work was financially supported by the Hong Kong Research Grant Council (GRF Project No. 16206119) and the Fundamental Research Program of Shenzhen (Grant No. JCYJ20170412153039309). The authors acknowledge the assistance of Advanced Engineering Materials Facility (AEMF) of HKUST and SUSTech Core Research Facilities. The authors acknowledge Kangjie Chu, Zhongzheng Deng and Qiuhong Wang for the kind assistance in TEM analysis and mechanical testing.

# Functional Degradation and Self-enhanced Elastocaloric Cooling Performance of NiTi Tubes under Cyclic Compression

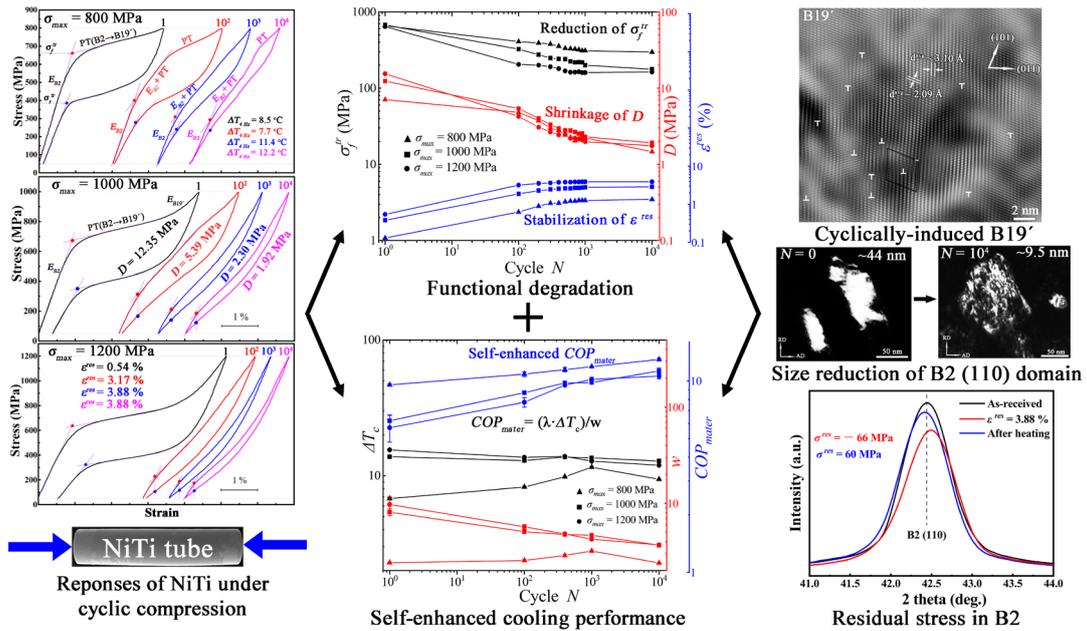

Graphical abstract

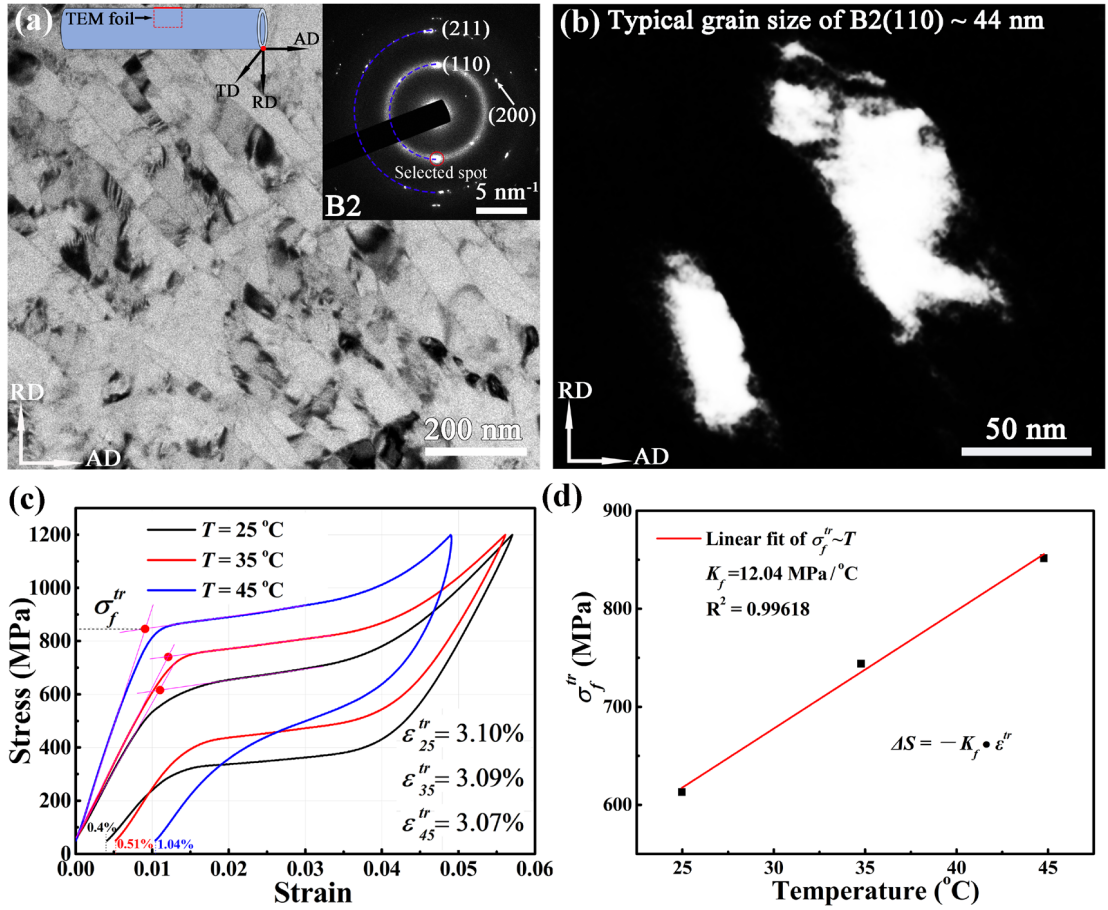

**Fig. 1.** (a) Bright field TEM image of as received NiTi tube with radial direction (RD) and axial direction (AD) labelled; (b) dark field TEM image taken with the diffraction spot of B2 (110); (c) isothermal compressive σ~ε responses of NiTi tubes at 25, 35 and 45 °C; (d) temperature dependence of the forward transformation stress $\sigma_f^{tr}$.

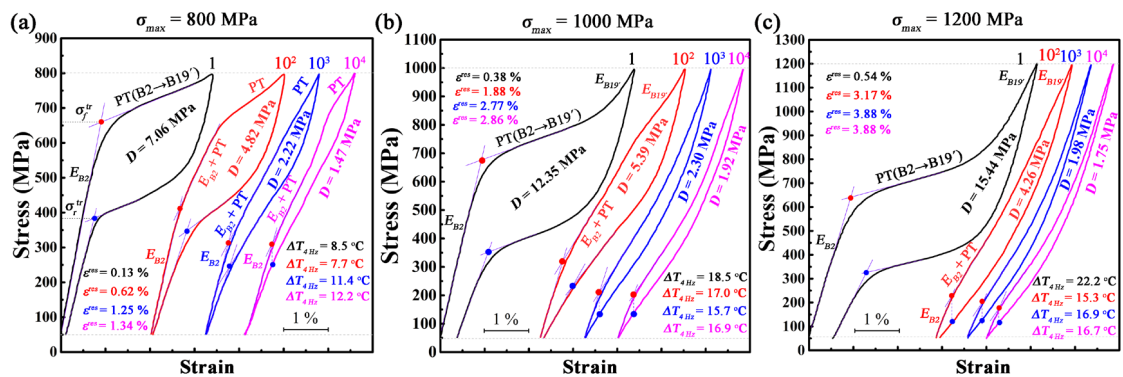

**Fig. 2.** Evolution of isothermal compressive σ~ε responses under (a) $\sigma_{max}$ = 800 MPa, (b) $\sigma_{max}$ =1000MPa and (c) $\sigma_{max}$ =1200 MPa.

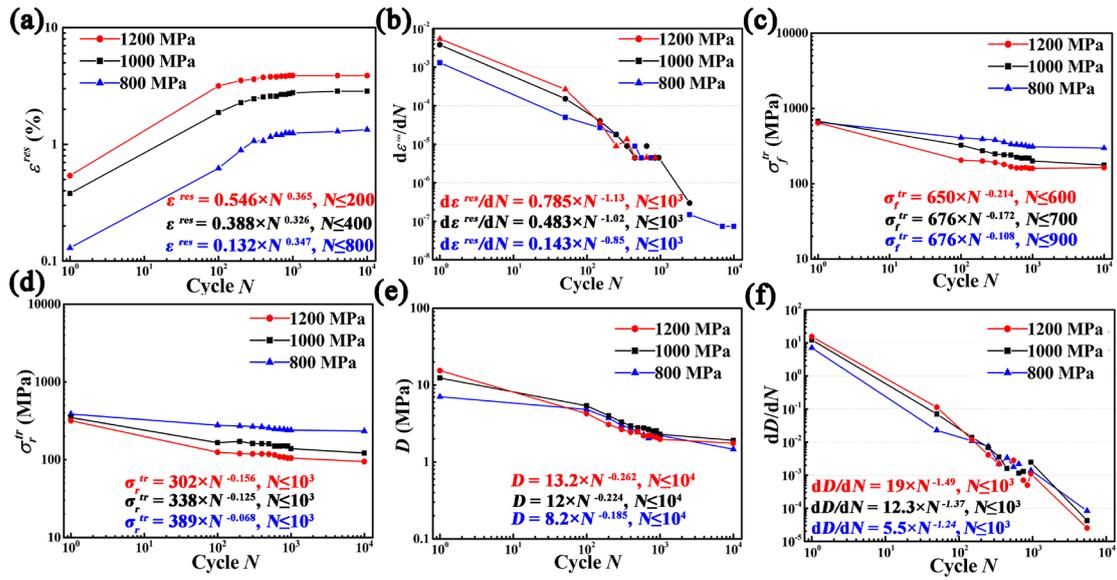

**Fig. 3.** Functional degradation of the NiTi tube under different stress levels as characterized by (a) $\varepsilon^{res}$~$N$ and (b) $d\varepsilon^{res}/dN$~$N$; (c) $\sigma_f^{tr}$~$N$ and (d) $\sigma_r^{tr}$~$N$; (e) $D$~$N$ and $dD/dN$~$N$.

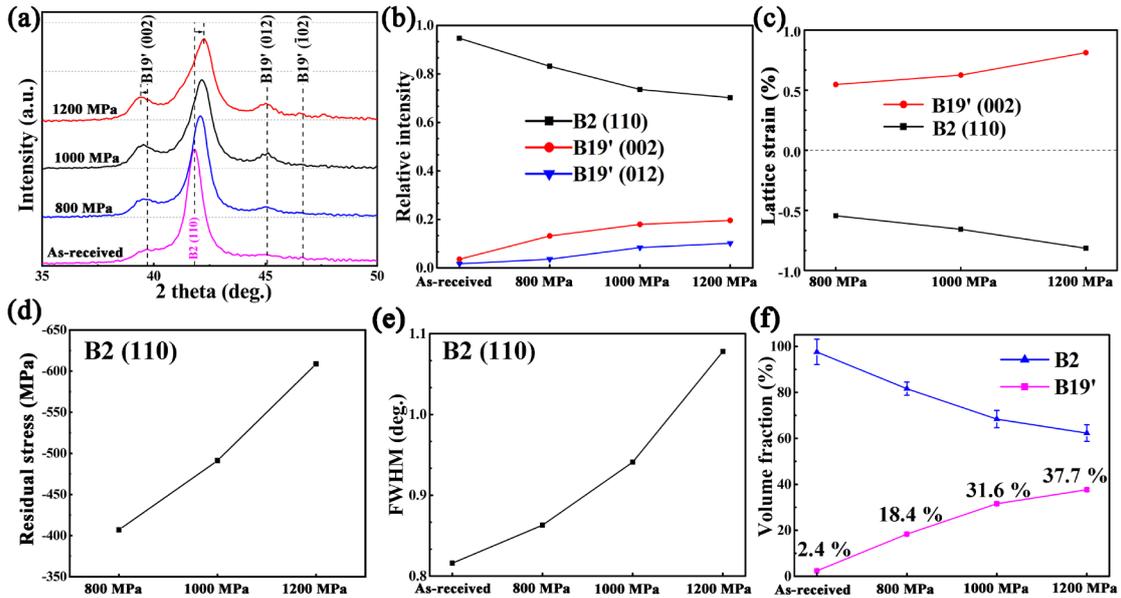

**Fig. 4.** (a) XRD patterns of cyclically-deformed tubes after $10^4$ cycles under 800 MPa, 1000 and 1200 MPa; (b) relative intensity of diffraction peaks of B2 and B19′; (c) lattice strain; (d) residual stress in austenite B2 (110) at $N=10^4$ versus $\sigma_{max}$; (e) FWHM of B2(110); (f) volume fraction of B2 and B19′ estimated via Rietveld refinement.

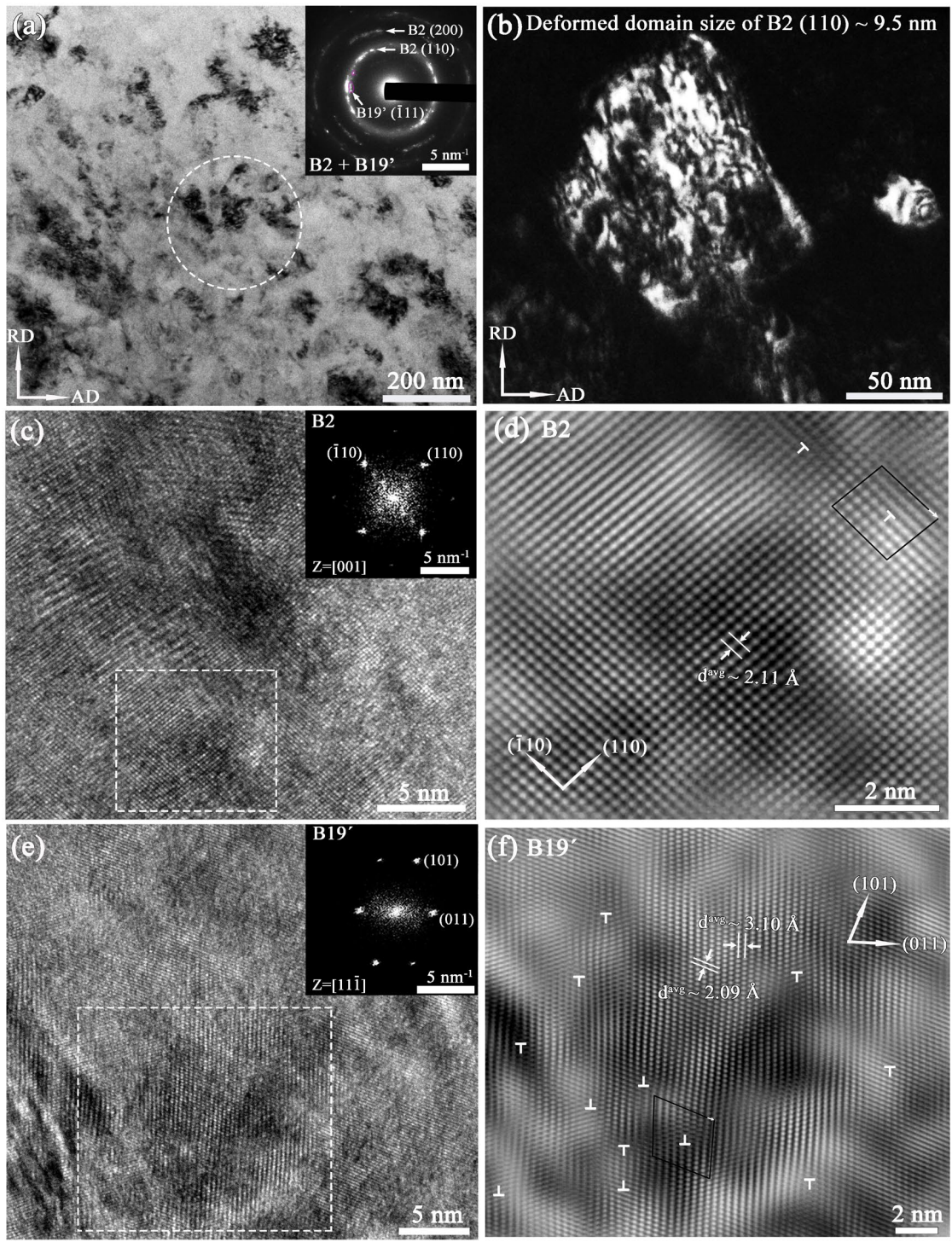

**Fig. 5.** Cyclically-deformed microstructure after $10^4$ cycles under 1200 MPa. (a) The bright field TEM image with SADP (inset); (b) the dark field TEM image taken with the diffraction spot of B2 (110); (c) the high resolution TEM image of B2 austenite; (d) the inverse fast Fourier transformation (IFFT) image from the selected area in (c); (e) the high resolution TEM image of B19′ martensite; (d) the IFFT image from the selected area in (e).

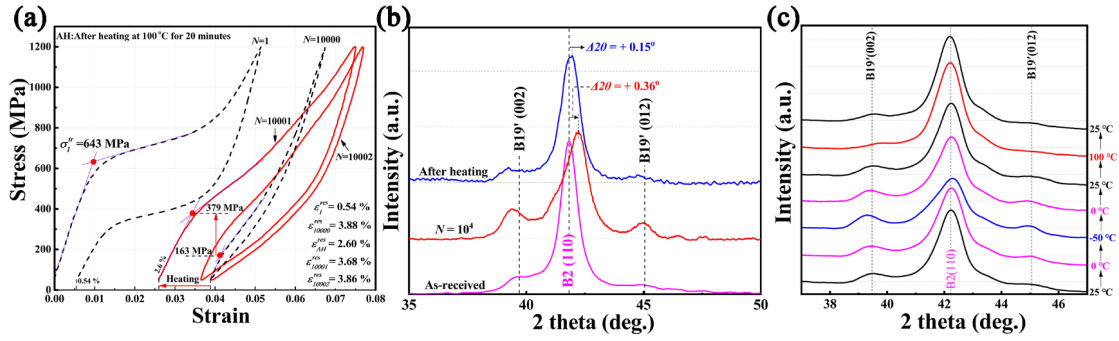

**Fig. 6.** (a) Comparison of isothermal σ~ε responses among $N = 1$, $N = 10^4$, $N = 10001$ and $N = 10002$ (tested after heating at 100 ºC for 20 minutes); (b) comparison of XRD patterns of the as-received tube, the tube after $10^4$ cycles of compression under $σ_{max}$=1200 MPa and the tube after heating at 100 ºC for 20 minutes; (c) *in situ* XRD patterns of the deformed tube ($N = 2 \times 10^4$, $ε^{res} = 3.88\%$).

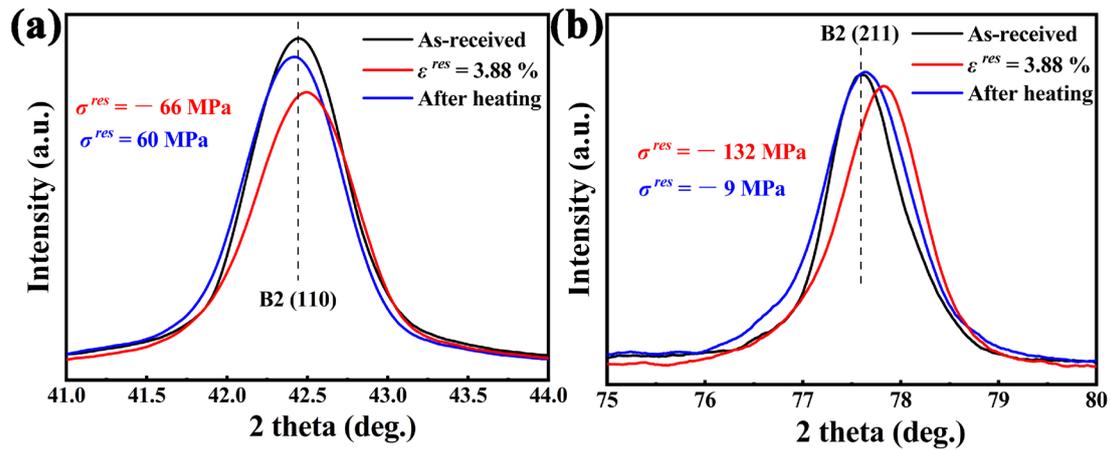

**Fig. 7.** XRD patterns of as-received tube, the deformed tube ($N = 2 \times 10^4$, $ε^{res} = 3.88\%$) and the deformed tube after heating along the loading direction. (a) B2 (110); (b) B2 (211)

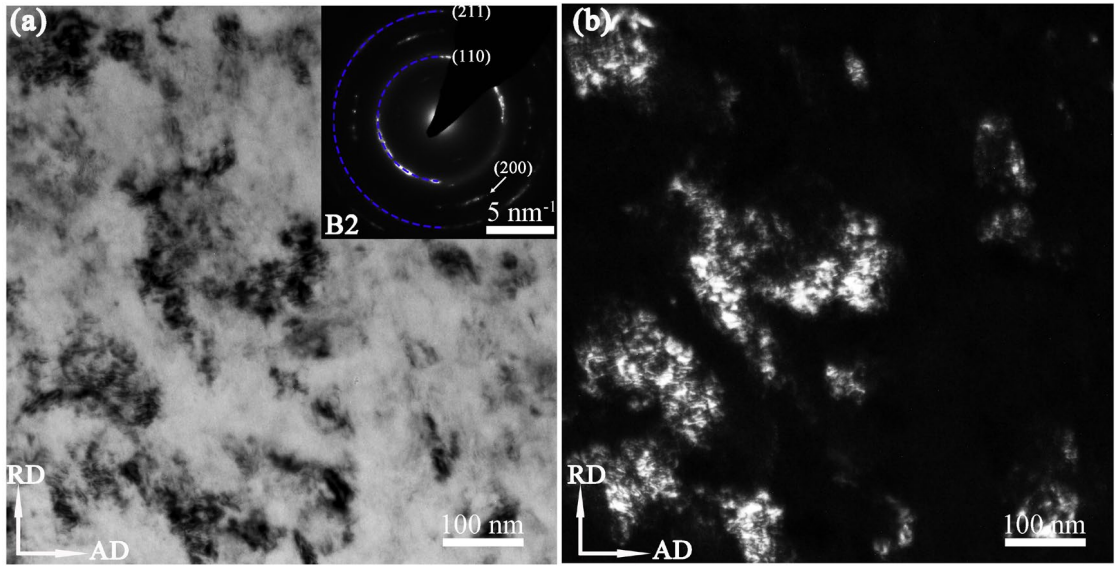

**Fig. 8.** Microstructure of the cyclically-deformed tube after heating at 100 °C for 20 minutes. (a) The bright field TEM image with the selected area diffraction pattern; (b) the dark field TEM image taken with the diffraction spot of B2 (110).

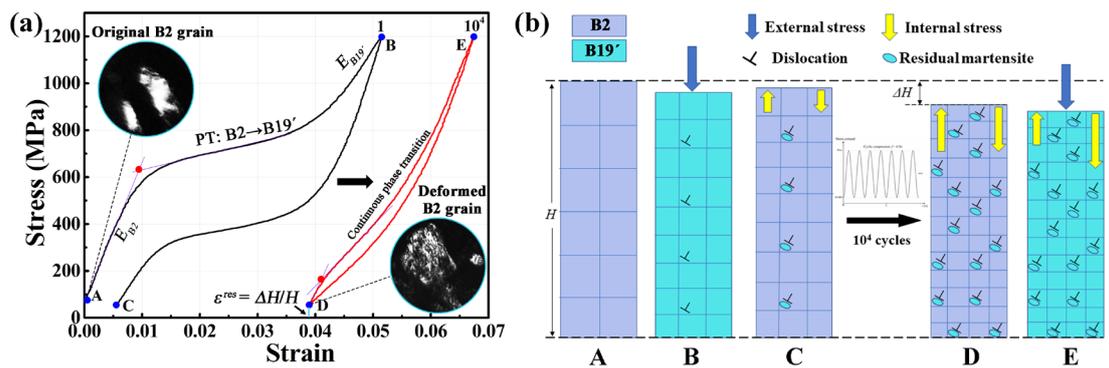

**Fig. 9** (a) Isothermal $\sigma\sim\varepsilon$ responses at $N=1$ and $N=10^4$; (b) schematic illustration of the stabilization mechanism of the NiTi tube under cyclic compression.

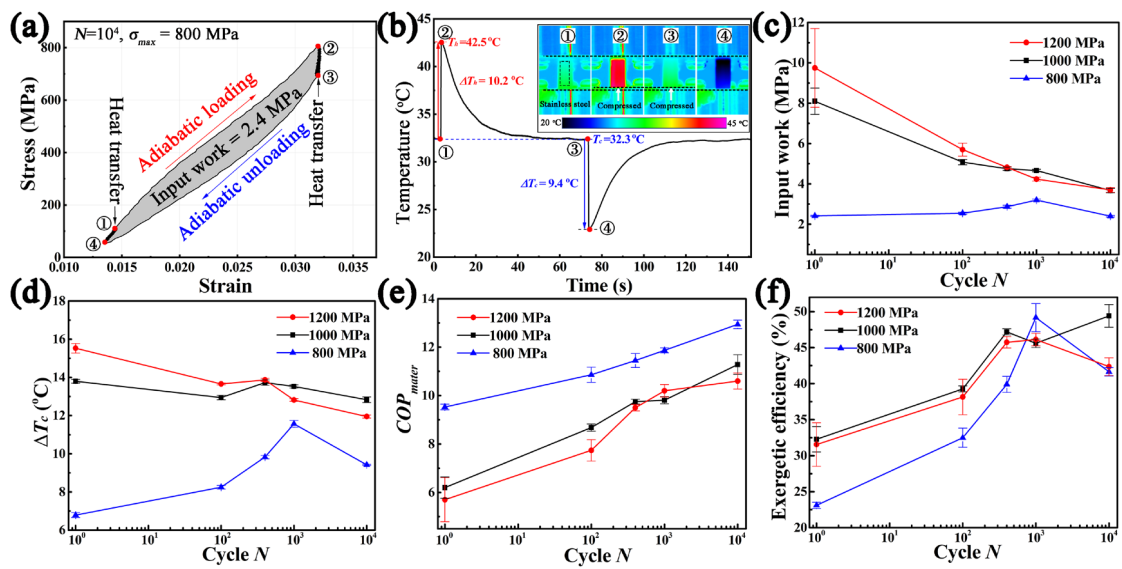

**Fig. 10** Self-enhancement of elastocaloric cooling under different stress levels. (a) the mechanical $\sigma\sim\varepsilon$ response of *COP* measurement under $\sigma_{max} = 800$ MPa at $N=10^4$; (b) *in situ* detected temperature profile; evolution of input work (c), adiabatic cooling temperature drop (d), $COP_{mater}$ (e) and exergetic efficiency (d).

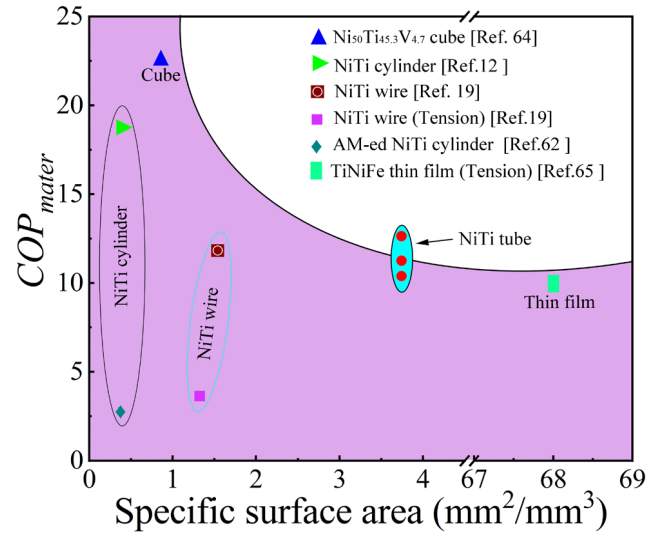

**Fig. 11** Comparison of specific surface area and $COP_{mater}$ for NiTi tube and other NiTi based SMAs.

# Supplementary Materials

# Functional Degradation and Self-enhanced Elastocaloric Cooling Performance of NiTi Tubes under Cyclic Compression

Dingshan Liang[1,2], Peng Hua[1], Junyu Chen[3], Fuzeng Ren[2,*], Qingping Sun[1,*]

[1] Department of Mechanical and Aerospace Engineering, The Hong Kong University of Science and Technology, Clear Water Bay, Kowloon, Hong Kong, China

[2] Department of Materials Science and Engineering, Southern University of Science and Technology, Shenzhen, Guangdong, China

[3] Department of Engineering Mechanics, School of Civil Engineering, Wuhan University, Wuhan, Hubei, China

Corresponding Author. E-mail: renfz@sustech.edu.cn; meqpsun@ust.hk

# 1. Determination of austenite finish temperature

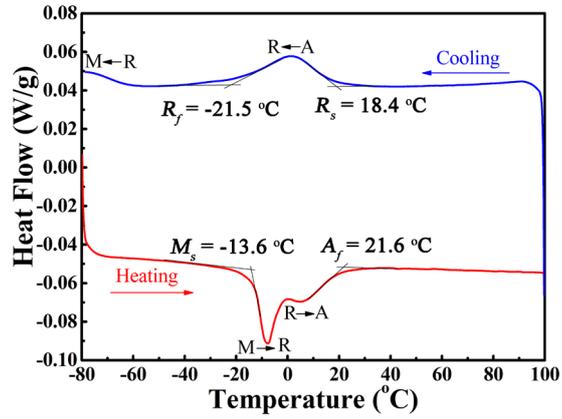

**Fig. S1.** Differential scanning calorimetry curves of the NiTi tube ranging from −80 ºC to 100 ºC.

# 2. Experimental set-up

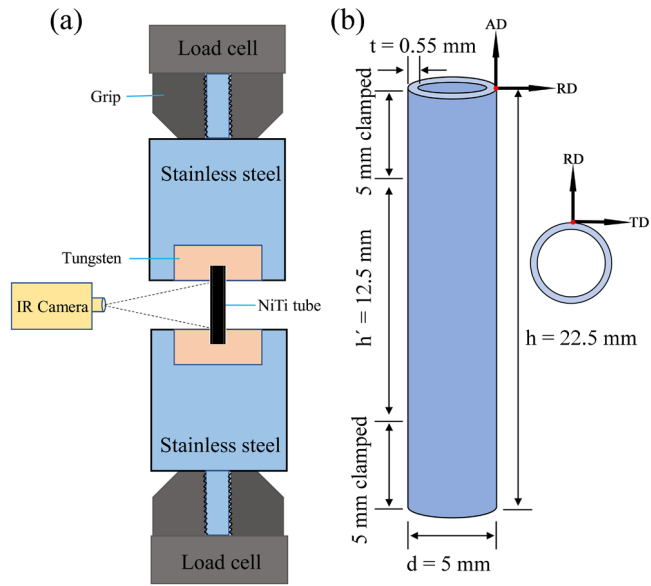

**Fig. S2.** (a) Schematic illustration of the designed clamping compression fixture for NiTi tubes; (b) Geometrical dimensions of NiTi tubes with an effective height (h′) of 12.5 mm. The radial direction, axial direction and tangential direction of NiTi tubes were denoted as RD, AD and TD, respectively.

# 3. Estimation of buckling

The elastic buckling of solid structure subjected to clamping conditions can be estimated by

$$P_{cr} = \frac{4\pi^2 EI}{l^2} \quad (1)$$

where $P_{cr}$ is the critical load, $E$ is the elastic modulus, $I$ is the second momentum of inertia, $l$ is the length [1]. However, the elastic modulus of superelastic NiTi is highly temperature-sensitive and the slope of strain-stress curve during phase transformation region is very small. That is, the NiTi tube might not be buckled at the elastic region but easily buckled at the PT region. To be more conservative, take the slope of PT region during isothermal strain-stress curve at 25 °C as $E_{PT}$, substituting the elastic modulus in equation (1). For NiTi tube with outer diameter of 5 $mm$ and inner diameter of 3.9 $mm$: the maximum applied stress $P_{cr}$ is 1200 MPa; $E_{PT}$ is 4285 MPa from Fig. 1(c) ; $I$ is 19.32 mm⁴. So, the maximum allowed length $l_{max}$ is 18.819 $mm$. In the clamping head as showed in Fig. S2, the effective length of the tube is 12.5 $mm$, which is smaller than the $l_{max}$. Therefore, the tube will not buckle under such clamping condition.

### 4. Monoclinic compression test

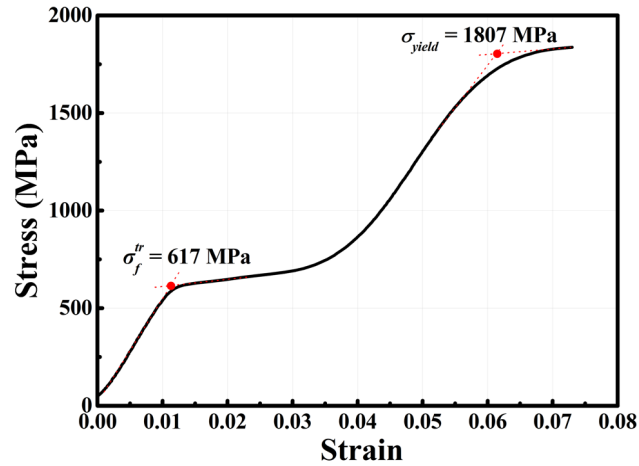

**Fig. S3.** Monotonic compression of NiTi tubes subjected to plastic deformation at room temperature.

### 5. Temperature drop under 800 MPa

As shown in Fig. S4a, the tube was loaded to 800 MPa slowly and held the force to attain a maximum strain, then adiabatically unloaded in 0.2 S to get a temperature drop.

Fig. S4b shows the temperature was first increased to about 35 °C and then cooled back to near the level of room temperature. An adiabatic temperature drop of 25 °C was observed during the unloading process.

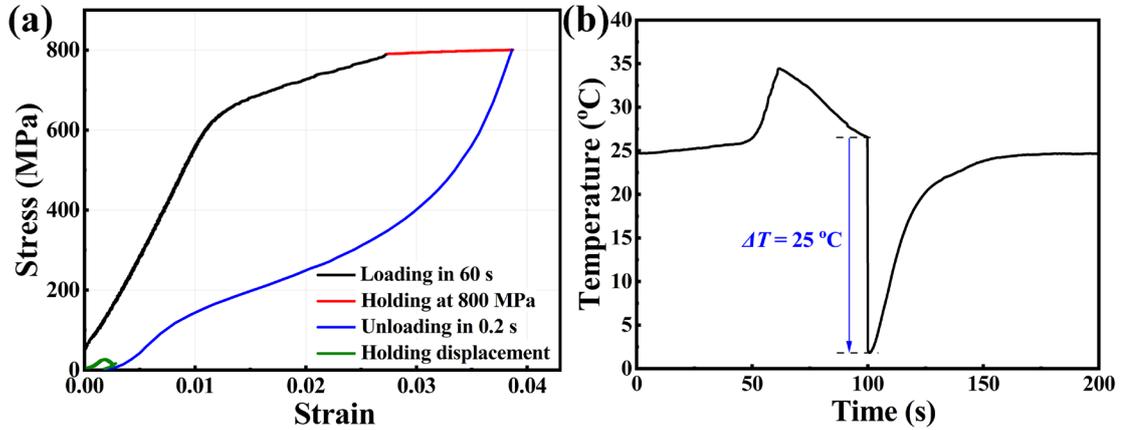

**Fig. S4.** (a) Mechanical stress-strain response of the measurement of maximum temperature drop; (b) *in situ* temperature profile with a temperature drop of 25 °C.

## 6. Evolution of 4 Hz cyclic compression

In contrast to the isothermal response tested after certain cycles of compression, the cyclic compression at 4 Hz was conducted to accumulate cycle number $N$ continuously within a period. The first, $100^{th}$, $1000^{th}$ and $10000^{th}$ cycle in the cyclic compression with *in situ* detected temperature profile were presented in Fig. S5 to characterize the evolutions coupling with heat generation (forward phase transition and hysteresis heat) and absorption (reverse phase transition ) under various $\sigma_{max}$.

Under $\sigma_{max}$ of 800 MPa, the shape of strain-stress curve involves from a top-right broad "leaf" with residual strain at the beginning to a slim "needle" sharping at both end after $10^4$ cycles (Fig. S5a). As a result, the amplitude of temperature oscillation ($\Delta T$) first drops from 8.5 °C to 7.7 °C for the first 100 cycles of compression, and then gradually improves to 11.4 °C at $N=10^3$ and 12.2 °C at $N=10^4$ (Fig. S5b). As the cycles accumulated, the faster the temperature rises versus time (Fig. S5c). These can be related to the decrease of $\sigma_f^{tr}$, so the temperature rises quickly at the low stress level. Not that the maximum stress was achieved at t = 0.125s.

Under $\sigma_{max}$ of 1000 MPa and 1200 MPa, the shape of $\sigma$-$\varepsilon$ curves both change from middle-wide open loops to slim sickle-like types with oblivious hardening in the up-right ends (Fig. S5d&g). The obvious hardening at the stage of elastic deformation is mainly due to the preferred orientation of stress-induced martensite along the loading direction possessing high modulus [2]. The hysteresis loop area shrinks first from 4.18 MPa at the beginning to 1.72 MPa at $N$=100 and stabilizes to 1.63 MPa after $10^3$ cycles under $\sigma_{max}$ = 1000 MPa. While it keeps decreasing non-monotonically from 7.47 MPa at the beginning to 1.49 MPa at $N$=$10^4$ for the case under $\sigma_{max}$ of 1200 MPa. The reversible strain behaves like $\Delta T$ under $\sigma_{max}$ = 1000 MPa, both demonstrating a reduction first from the beginning to $N$=100 and then a gradual increase till $N$=$10^4$. The time dependence of temperature is similar as that under $\sigma_{max}$ = 800 MPa, showing the more cycles compressed the quicker temperature rises. Under $\sigma_{max}$ = 1200 MPa, the reversible strain manifests a rapid reduction from 3.4% to 2.6% during the first 100 cycles and then evolves very slowly to 2.5% and 2.4% for $N$ = $10^3$ and $N$=$10^4$ (Fig. S5e). As for the temperature oscillation, it first drops from 22.2 °C to 15.3 °C for the first 100 cycles, then rises to about 16.9 °C at $N$ = $10^3$, and final decreases slowly (Fig. S5i). The time dependence of temperature is not obvious under $\sigma_{max}$ = 1200 MPa. This might be ascribed to the coupling of latent heat release and hysteresis heat. For the first cycle, it is notable that the temperature at $t$ = 0.25 s cannot fall back to the same level at $t$ = 0 s, displaying a gap proportional to the maximum applied stress(Fig. S5b,e&h). This should be accredited to the continuous release of hysteresis heat during unloading.

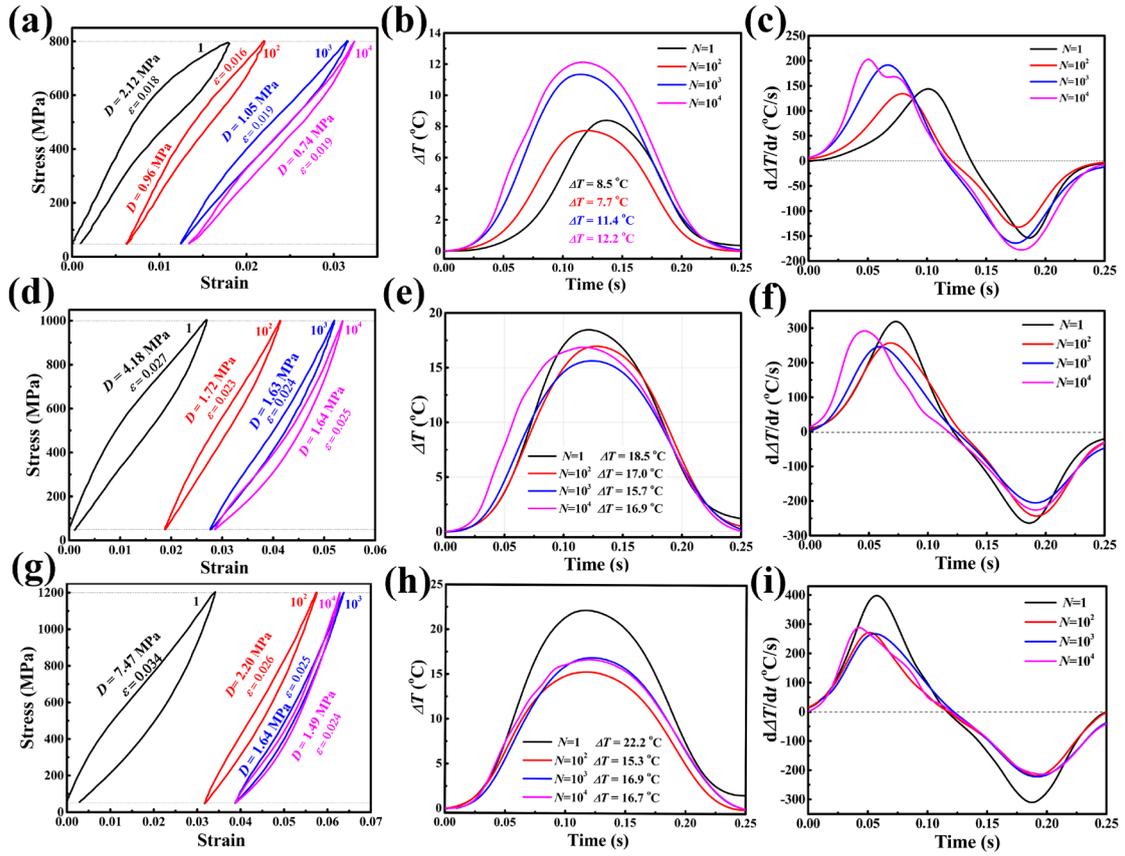

**Fig. S5.** Mechanical evolution of 4 Hz cyclic compression under (a) $\sigma_{max}$ = 800 MPa, (d) $\sigma_{max}$ = 1000MPa and (g) $\sigma_{max}$ = 1200 MPa; *In situ* temperature profile as a function of time under (b) $\sigma_{max}$ = 800 MPa, (e) $\sigma_{max}$ =1000MPa and (h) $\sigma_{max}$ =1200 MPa; d$\Delta T$/dt as a function of time under (c) $\sigma_{max}$ = 800 MPa, (f) $\sigma_{max}$ =1000MPa and (i) $\sigma_{max}$ =1200 MPa.

## 7. Rietveld refinement via MAUD

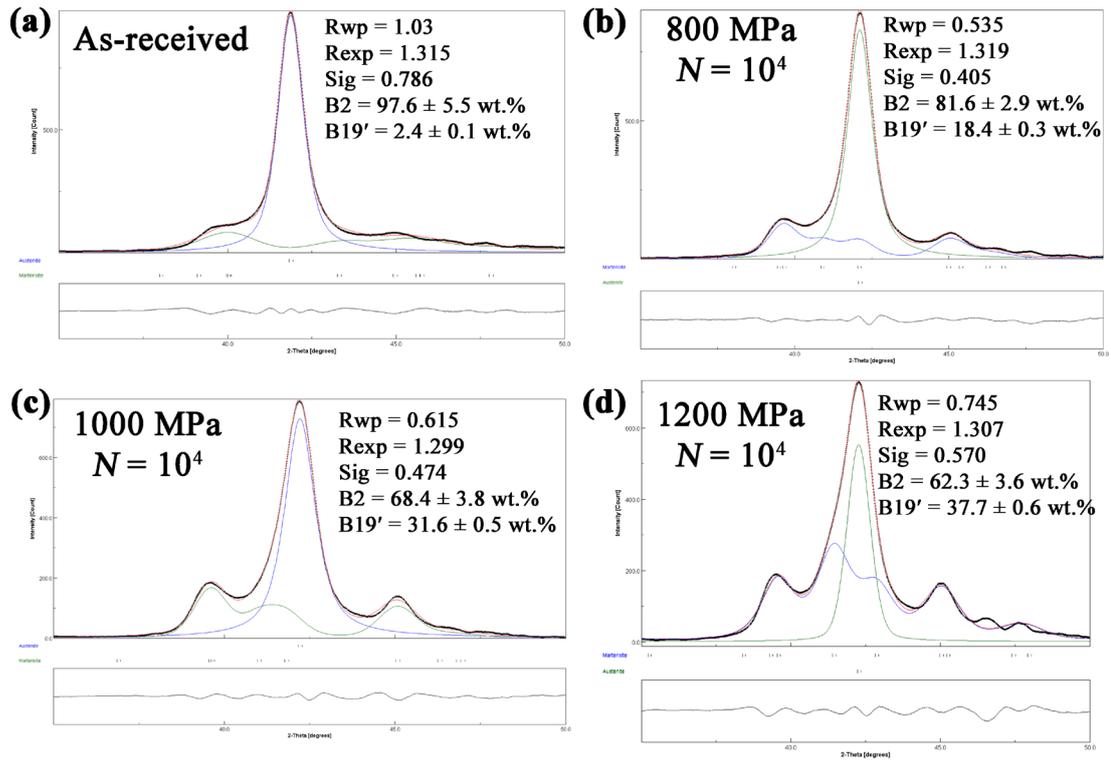

**Fig. S6.** Rietveld refinement analysis of the XRD patterns obtained from the tube surface of (a) as-received, cyclically-deformed under 800 MPa (b), 1000 MPa (c) and 1200 MPa (d) tubes.

## 8. Surface morphologies evolution

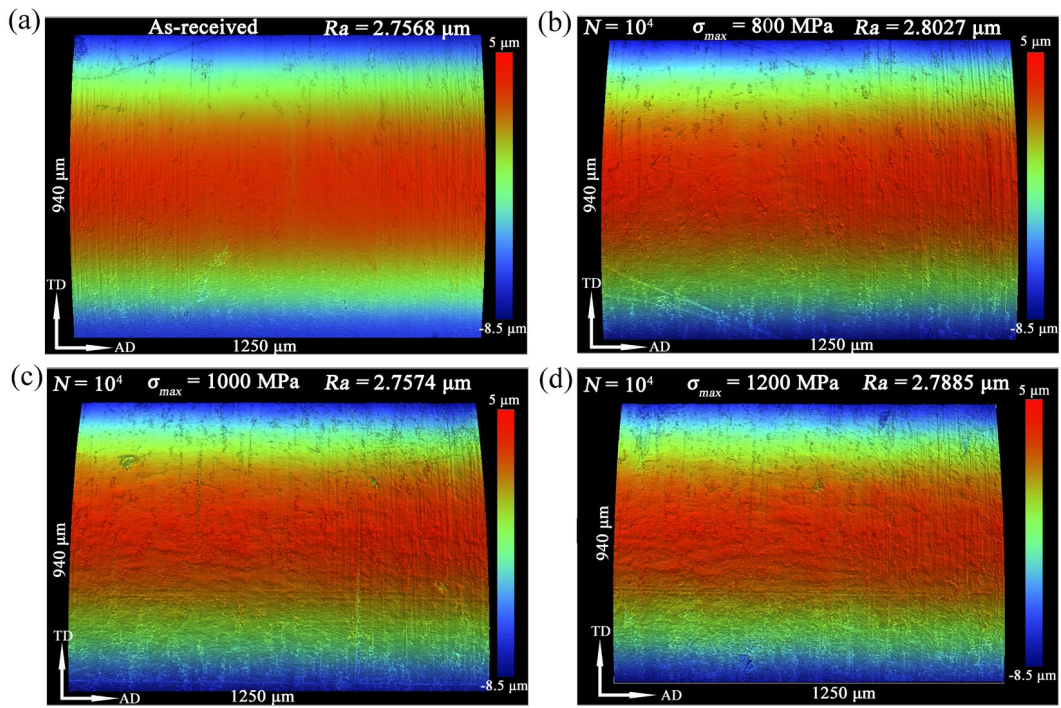

**Fig. S7.** 3D surface profiles of as-received tubes and cyclically-deformed tubes.

The increased surface roughness (*Ra*) of the cyclically-deformed tubes should be a result of phase-transformation induced plasticity.

## 9. Residual lattice strain in B2 via SAED

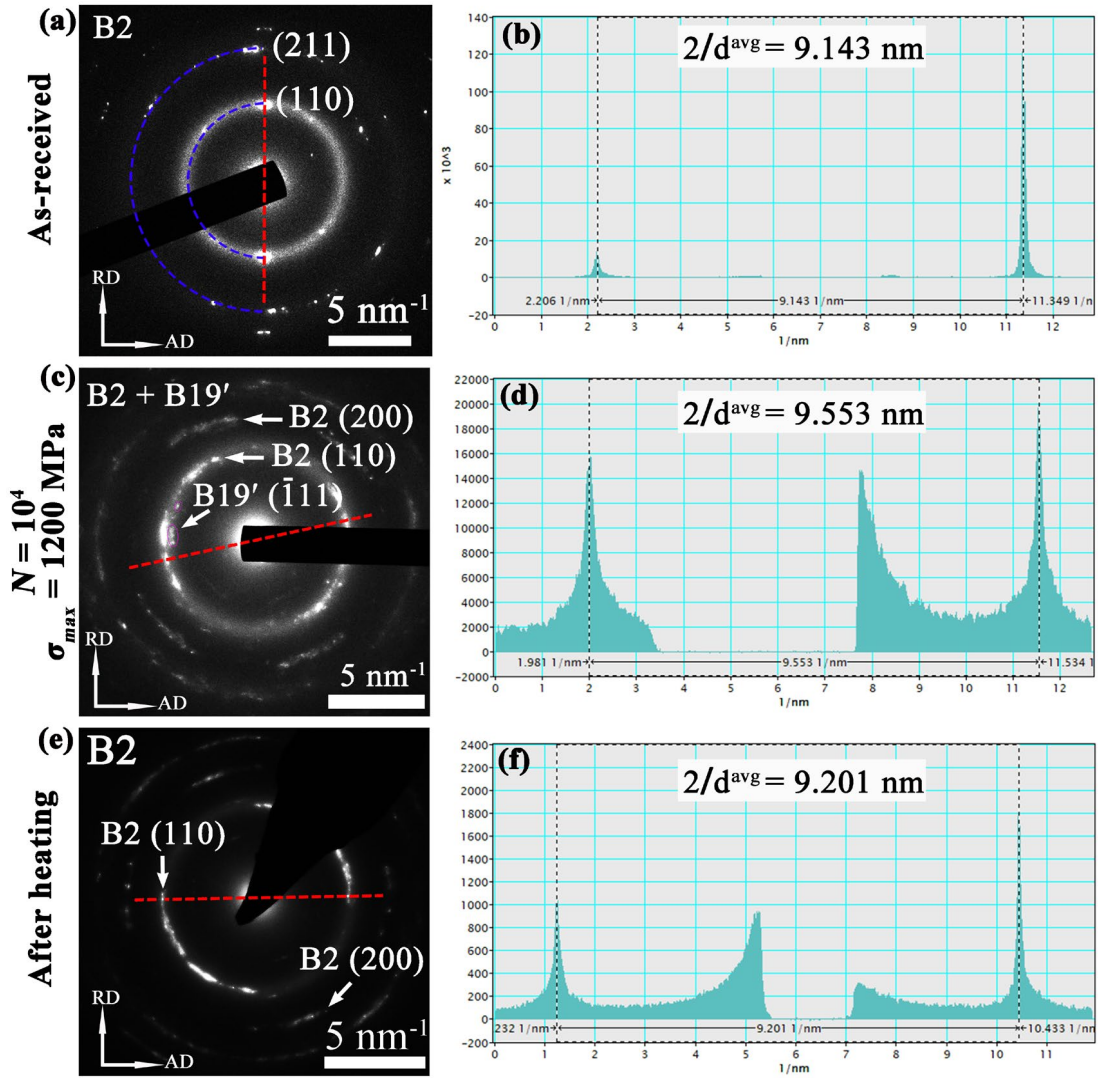

**Fig. S8.** SADP of as-received tube (a), deformed tube (c) and the deformed tube after heating (e); (b), (d) and (f) corresponding line profile of the red line in (a), (c) and (e).

## 10. Relation between hysteresis loop area and difference of $\Delta\sigma^{tr}$

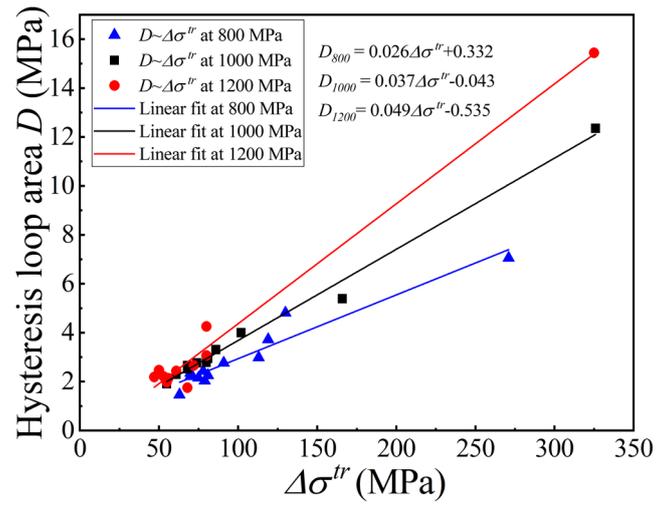

**Fig. S9.** Relationship of hysteresis loop area and difference of $\sigma_f^{tr}$ and $\sigma_r^{tr}$.

## 11. The *COP* measurements at *N*=1 and *N*=10⁴

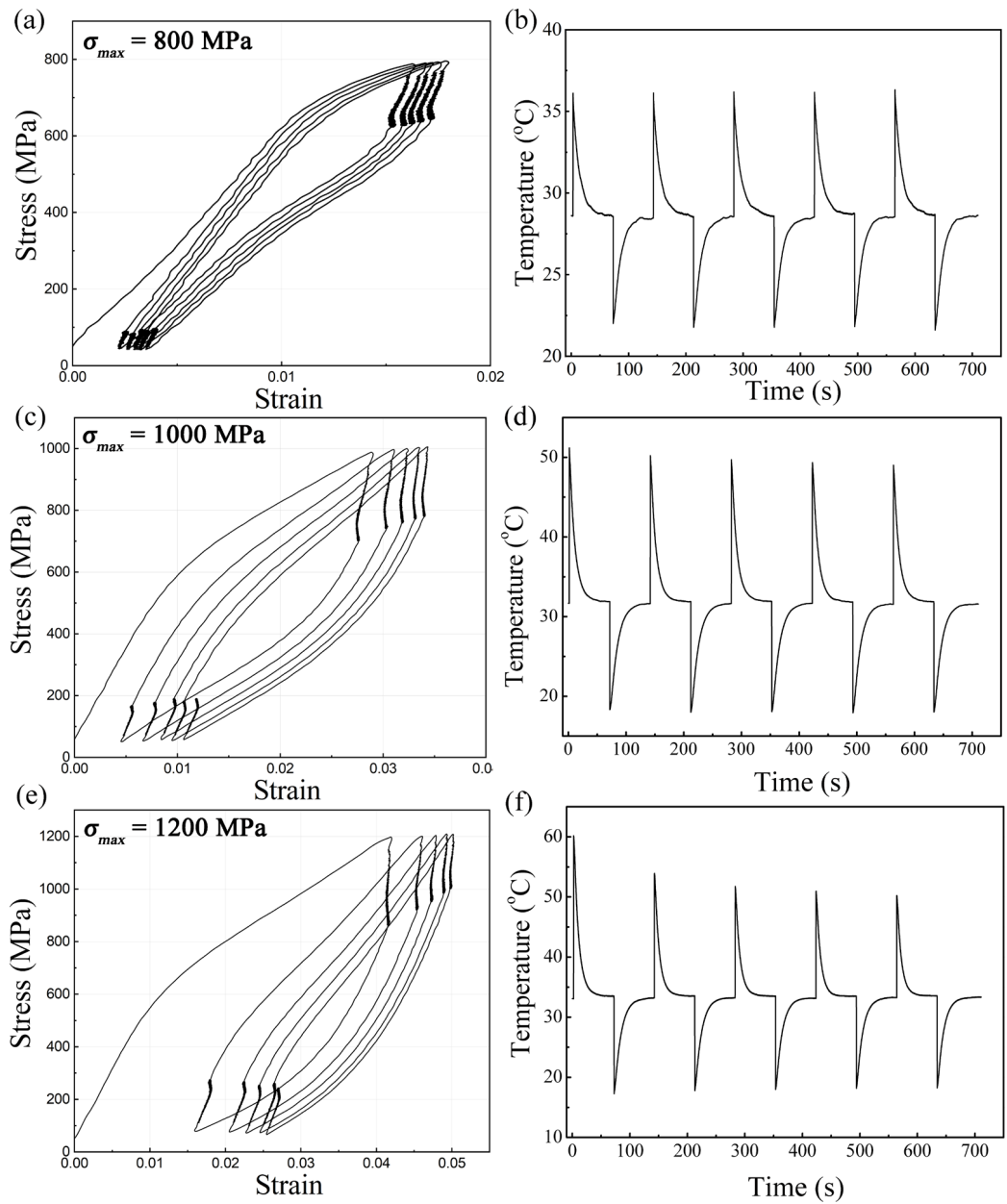

**Fig. S10.** Mechanical response at *N* = 1 for COP measurement under $\sigma_{max}$ of 800 MPa (a), 1000MPa (c) and 1200 MPa (e); *In situ* temperature profile as a function of time under $\sigma_{max}$ of 800 MPa (b), 1000MPa (d) and 1200 MPa (f).

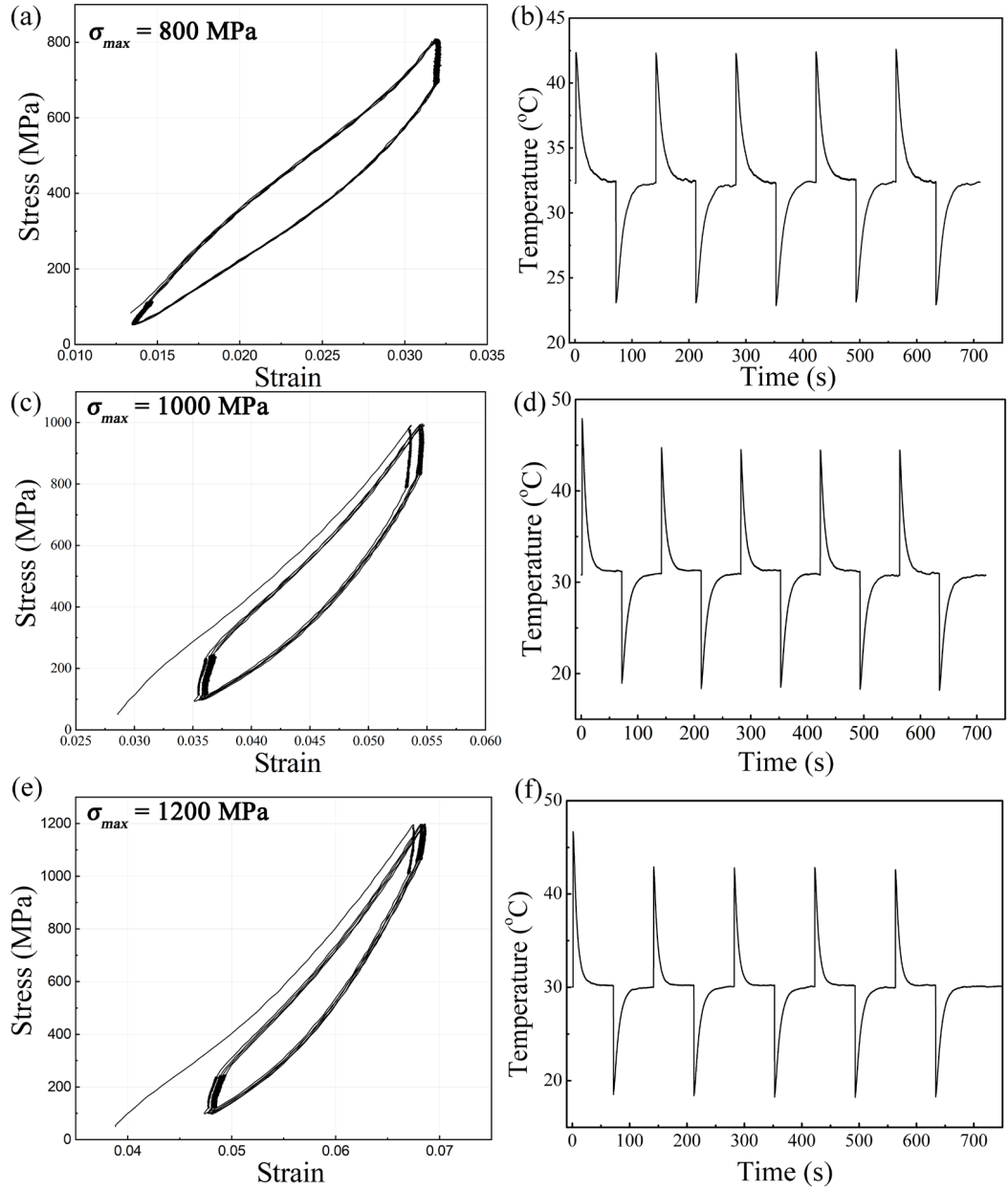

**Fig. S11.** Mechanical response at $N = 10^4$ for COP measurement under $\sigma_{max}$ of 800 MPa (a), 1000MPa (c) and 1200 MPa (e); *In situ* temperature profile as a function of time under $\sigma_{max}$ of 800 MPa (b), 1000 MPa (d) and 1200 MPa (f).

**12. Lattice strain estimated by geometric phase analysis**

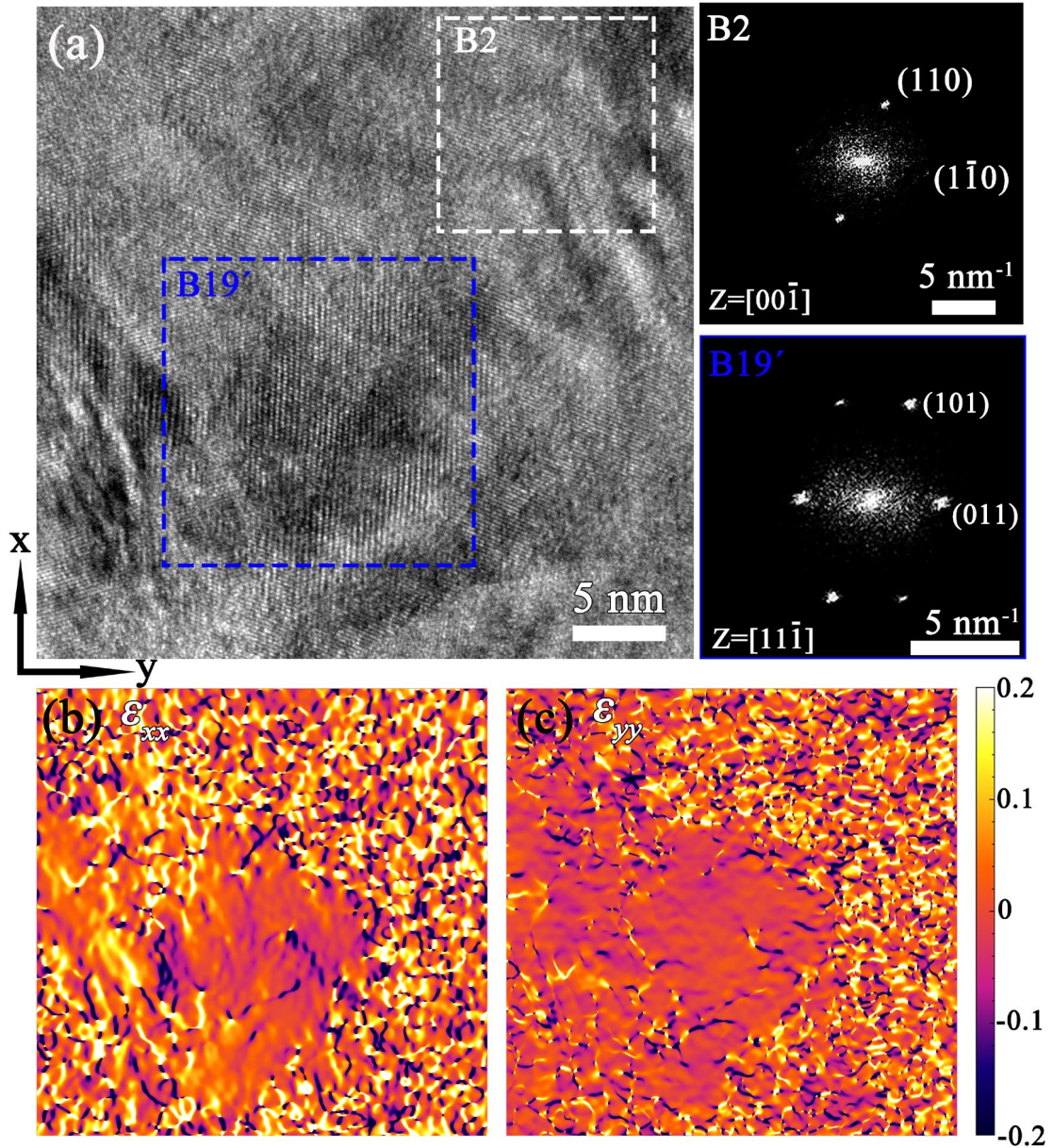

**Fig. S12.** Lattice strain estimated by geometric phase analysis via Strain++ [3] after $10^4$ under 1200 MPa. (a) High resolution TEM of B2 and B19′; (b) estimated $\varepsilon_{xx}$; (c) estimated $\varepsilon_{yy}$. The strain field in B2 is in a chaos of compressive and tensile strain, randomly distributing at the sub-nano scale. However, it is basically tensile strain in B19′, despite highly localized compressive strain at the phase boundaries. Such qualitive estimation is consistent to the discussions of the main text.